\documentclass[12pt,english]{article}
 \usepackage{helvet}
 \usepackage[T1]{fontenc}
 \usepackage[cp1251]{inputenc}
 \usepackage{geometry}
 \usepackage{graphicx}
 \usepackage{setspace}
 \usepackage{babel}

 \renewcommand{\vec}[1]{\mbox{\boldmath $#1$}}

 \def\gsim{\lower.4ex\hbox{$\;\buildrel >\over{\scriptstyle\sim}\;$}}
 \def\lsim{\lower.4ex\hbox{$\;\buildrel <\over{\scriptstyle\sim}\;$}}
 \def\newpage{\vfill\eject}
 \def\bl{\par\vskip 12pt\noindent}
 \def\bll{\par\vskip 24pt\noindent}
 \def\blll{\par\vskip 36pt\noindent}

 \def\beg{\begin{eqnarray}}
 \def\ende{\end{eqnarray}}

 \def\araa{ARA\&A}
 \def\aa{A\&A}
 \def\apj{ApJ}
 \def\apjs{ApJS}
 \def\an{Astron. Nachr.}
 
 \def\mnras{MNRAS}

 \def\aj{Astron. Rep.}

\begin{document}

\begin{center}
{\bf BAROCLINIC INSTABILITY} \\[0.2 truecm]
{\bf IN DIFFERENTIALLY ROTATING STARS}
\end{center}

\bll

\centerline{L.\,L.~Kitchatinov$^{1,2}$\footnote{E-mail:
kit@iszf.irk.ru}}

\bl

\begin{center}
$^1${\it Institute for Solar–Terrestrial Physics, P.O. Box 4026, Irkutsk, 664033 Russia} \\
$^2${\it Pulkovo Astronomical Observatory, Pulkovskoe Sh. 65, St. Petersburg, 196140 Russia}
\end{center}

\bll
\hspace{0.8 truecm}
\parbox{14.4 truecm}{
{\bf Abstract.} A linear analysis of baroclinic instability in a stellar radiation zone with radial differential rotation is performed. The instability onsets at a very small rotation inhomogeneity, $\Delta\Omega \sim 10^{-3}\Omega$. There are two families of unstable disturbances corresponding to Rossby waves and internal gravity waves. The instability is dynamical: its growth time of several thousand rotation periods is short compared to the stellar evolution time. A decrease in thermal conductivity amplifies the instability. Unstable disturbances possess kinetic helicity thus indicating the possibility of magnetic field generation by the turbulence resulting from the instability.
 }

\bll

{\bf DOI:} 10.1134/S1063773713080045

\bll

Keywords: {\sl stars: rotation – instabilities – waves.}

\blll

\reversemarginpar

\setlength{\baselineskip}{0.8 truecm}

%%%%%%%%%%%%%%%%%%%%%%%%%%%%%%%%%%%%%%%%%%%%%%%%%%%%%%%%%%%%%%%%%%%
 \centerline{\bf INTRODUCTION}
 \bl
%%%%%%%%%%%%%%%%%%%%%%%%%%%%%%%%%%%%%%%%%%%%%%%%%%%%%%%%%%%%%%%%%%%
Upon arrival on the main sequence, young stars
rotate rapidly, with periods of about one day. Solar-type
stars spin down with age due to the loss of
angular momentum through a stellar wind (Skumanich
1972; Barnes 2003). The braking torque acts
on the stellar surface, but the spin-down extends
rapidly deep into the convective envelope due to the
eddy viscosity existing here. In deeper layers of the
radiation zone, the viscosity is low ($\sim 10$\,cm$^2$/s)
and insufficient to smooth out the radial rotation inhomogeneity.
Therefore, before the advent of helioseismology,
it had been thought very likely that the solar
radiation zone rotates much faster than the surface
(see, e.g., Dicke 1970). Subsequently, it transpired
that the radiation zone rotates nearly uniformly (Shou
et al. 1998). In other words, there is a coupling
between the convective envelope and deep layers of
the radiation zone that is efficient enough to smooth out
the rotation inhomogeneity in a short time compared
to the Sun’s age. Observations of stellar rotation
show that the characteristic time of the coupling is
$\lsim 10^8$\,yr (Hartmann \& Noyes 1987; Denissenkov
et al. 2010).

One possible explanation for the smoothing of
rotation inhomogeneities in stars is the instability of
differential rotation: the turbulence resulting from the
instability transports angular momentum in such a
way that the rotation approaches uniformity. The
difficulty of this explanation stems from the fact that
the threshold rotation inhomogeneity
 \begin{equation}
    q = -\frac{r}{\Omega}\frac{\mathrm{d}\Omega}{\mathrm{d}r}
    \label{1}
 \end{equation}
for the appearance of hydrodynamic instabilities is
not small, $q = O(0.1)$ (here, $\Omega$ is the angular velocity,
$r$ is the radius). One might expect that such
instabilities could reduce the rotation inhomogeneity
to its threshold value but could not remove it completely.
The baroclinic instability that is related to the rotation
inhomogeneity indirectly may constitute an exception (Tassoul \&
Tassoul 1983).
In the equilibrium state of a differentially rotating
radiation zone, the surfaces of constant pressure
and constant density do not coincide. Such a
“baroclinic” equilibrium can be unstable (Spruit \&
Knobloch 1984). In this paper, we consider the baroclinic
instability in a stellar radiation zone with radial
differential rotation. As we will see, the instability
occurs at a very small rotation inhomogeneity, $q \ll 1$.

Instabilities are also important for the mixing
of chemical species in stars (see, e.g., Pinsonneault
1997). The study of the stability of differential
rotation in stellar radiation zones has a long history
(Goldreich \& Schubert 1967; Acheson 1978; Spruit
\& Knobloch 1984; Korycansky 1991). This paper
differs in that we consider the stability against global
disturbances. The horizontal disturbance scale is
not assumed to be small compared to the stellar radius. At the same time, a stable stratification
of the radiation zone rules out mixing on a
large radial scale. Therefore, the radial disturbance
scale is assumed to be small. Such an approach
was applied to analyze the stability of latitudinal
differential rotation (Charbonneau et al. 1999; Gilman
et al. 2007; Kitchatinov 2010) in connection with
the problem of the solar tachocline. It showed that
the horizontally-global modes are actually
the dominant ones. As we will see, the same is
true for the baroclinic instability. The most unstable
disturbances correspond to global Rossby waves
($r$-modes) and internal gravity waves ($g$-modes),
which grow exponentially with time in the presence
of a radial rotation inhomogeneity. Therefore, the
baroclinic instability may be considered as the loss
of stability by a differentially rotating star with respect to the excitation of $r$- and $g$-modes
of global oscillations.

Both dominant modes possess kinetic helicity,
${\vec u}\cdot ( {\vec\nabla}\times{\vec u}) \neq 0$. Helicity is indicative the ability of a flow to generate magnetic fields (see, e.g., Vainshtein
et al. 1980). Here, our instability analysis joins with
another possible explanation for the uniform rotation
of the solar radiation zone, the magnetic field effect.
%%%%%%%%%%%%%%%%%%%%%%%%%%%%%%%%%%%%%%%%%%%%%%%%%%%%%%%%%%%%%%%%%%%
 \bll
 \centerline{\bf FORMULATION OF THE PROBLEM}
 \bl
%%%%%%%%%%%%%%%%%%%%%%%%%%%%%%%%%%%%%%%%%%%%%%%%%%%%%%%%%%%%%%%%%%%
 \centerline{\bf Background Equilibrium and the Origin of Instability}
 \bl
%%%%%%%%%%%%%%%%%%%%%%%%%%%%%%%%%%%%%%%%%%%%%%%%%%%%%%%%%%%%%%%%%%%
When analyzing the stability, we will assume the
initial equilibrium state to be stationary and cylinder symmetric about the rotation
axis. We consider the hydrodynamic stability, i.e., the
magnetic field is disregarded. We will proceed from
the stationary hydrodynamic equation
 \begin{equation}
    ({\vec V}\cdot{\vec\nabla}){\vec V} = -\frac{1}{\rho}{\vec\nabla}P - {\vec\nabla}\psi ,
    \label{2}
 \end{equation}
where $\psi$ is the gravitational potential, the standard
notation is used, the influence of viscosity on the
global flow is neglected. The main flow component
in the radiation zone is rotation,
 \begin{equation}
    {\vec V} = {\vec e}_\phi r \sin\theta \Omega ,
    \label{3}
 \end{equation}
where the usual spherical coordinates ($r,\theta ,\phi$) are used and ${\vec e}_\phi$ is the azimuthal unit vector. We assume the rotation to be sufficiently slow, $\Omega^2 \ll GM/R^3$, for the deviation of stratification from spherical
symmetry to be small. The stratification in stellar
radiation zones is stable, i.e., the specific entropy $s = c_\mathrm{v}\ln (P) - c_\mathrm{p} \ln (\rho )$, increases with radius $r$,
 \begin{equation}
    N^2 = \frac{g}{c_\mathrm{p}}\frac{\partial s}{\partial r} > 0.
    \label{4}
 \end{equation}

The buoyancy forces counteract the radial displacements.
Therefore, the meridional circulation is
small and the main flow component is rotation (3).
The characteristic time of meridional circulation in
the radiation zone exceeds the Sun’s age (Tassoul
1982). Nevertheless, the most important force
balance condition follows
from the equation for a meridional flow. This
condition can be derived by calculating the azimuthal
component of the curl of the equation of motion (\ref{2}).
This gives
 \begin{equation}
    r \sin\theta\frac{\partial\Omega^2}{\partial z} =
    -\frac{1}{\rho^2}\left({\vec\nabla}\rho\times{\vec\nabla}P\right)_\phi,
    \label{5}
 \end{equation}
where $\partial /\partial z = \cos\theta\partial /\partial r - r^{-1}\sin\theta\partial /\partial\theta$ is the spatial
derivative along the rotation axis, the subscript $\phi$ denotes the azimuthal component of the vector. The
centrifugal force is conservative only if the
angular velocity does not vary with distance $z$ from
the equatorial plane. The left part of Eq.\,(\ref{5}) allows for
the non-conservative part of the centrifugal force,
which by itself produces a vortical meridional flow. In
a stellar radiation zone, this non-conservative force is
balanced by the buoyancy force included in the right
part of Eq.\,(\ref{5}).

In the case of $z$-dependent differential rotation,
the equilibrium is baroclinic: the surfaces of constant
pressure and constant density do not coincide. One
might expect such an equilibrium to be unstable. This
can be seen after the following transformations of the
right part of Eq.\,(\ref{5}):
 \begin{equation}
    -\frac{1}{\rho^2}{\vec\nabla}\rho\times{\vec\nabla}P = \frac{1}{c_\mathrm{p}\rho}{\vec\nabla}s\times{\vec\nabla}P = \frac{1}{c_\mathrm{p}}{\vec\nabla}s\times{\vec g}^* ,
    \label{6}
 \end{equation}
where ${\vec g}^* = -{\vec\nabla}\psi + r\sin\theta\Omega{\vec e}_\phi\times{\vec\Omega}$ is the \lq\lq effective''
gravity. It can be seen from Eq. (6) that the isobaric
and isentropic surfaces do not coincide either. Figure~1 explains why an instability is possible in
this situation (Shibahashi 1980). For displacements
in the narrow cone between the isobaric and
isentropic surfaces, the gravitational forces increase the energy of the fluid particles being displaced. The relatively
light particles with a positive entropy (temperature)
perturbation are displaced opposite to the gravity, while the colder and relatively dense particles
are displaced in the direction of gravity. One
might expect the disturbances with such displacements
to be amplified due to the release of (gravitational)
energy of the equilibrium state. Remarkably,
the instability arises from the buoyancy forces that
usually exhibit a stabilizing effect in stellar radiation
zones.

\begin{figure}[htb]
 \centerline{
 \includegraphics[width=12 cm]{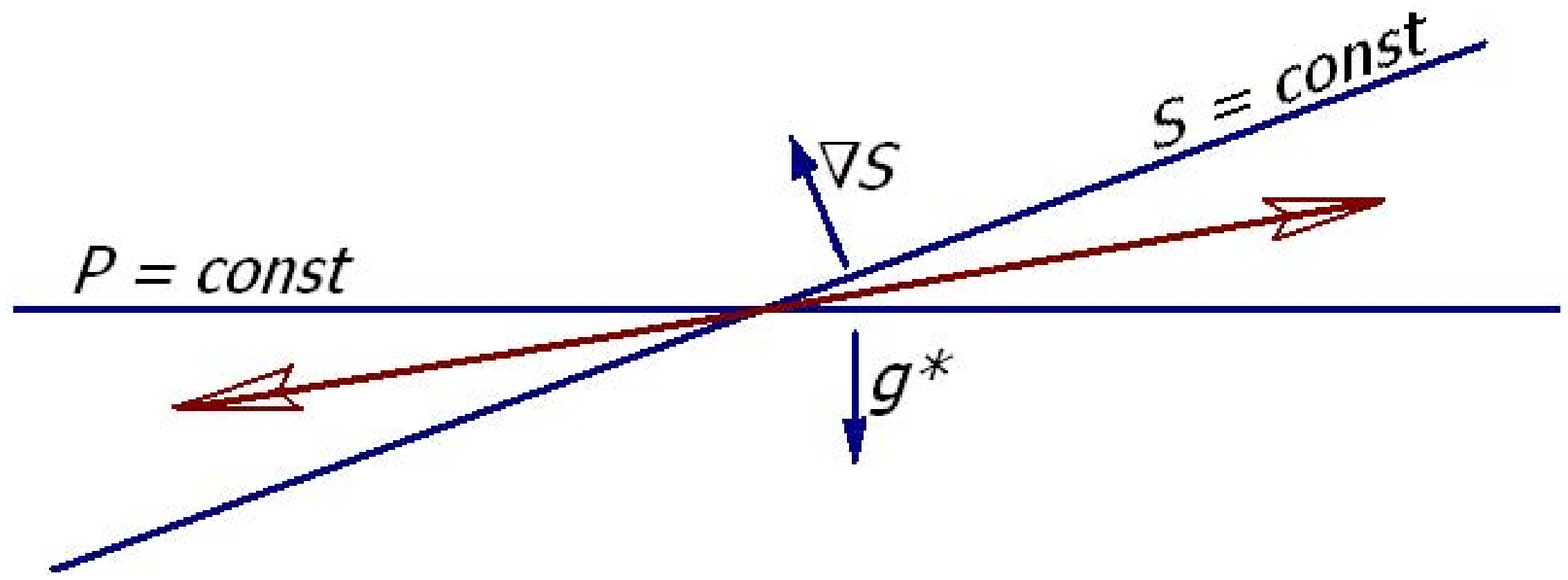}}
 \begin{description}
 \item{\small Fig.~1. If the isobaric and isentropic surfaces do not
    coincide, then the displacements
    in the cone between these surfaces (indicated
    by the arrows) can be unstable.
    }
 \end{description}
\end{figure}

It can be seen from Fig.~1 that not the deviation
of stratification from spherical symmetry related to
rotation but the baroclinicity caused by the rotation
inhomogeneity is responsible for the instability. For
simplicity, we will neglect the deviation of the pressure
distribution from spherical symmetry but will take
into account the latitudinal entropy inhomogeneity.
For the special case of rotation dependent only on the
radius, from Eqs. (\ref{5}) and (\ref{6}) we find
 \begin{equation}
    \frac{\partial s}{\partial\theta} = -2 q c_\mathrm{p} r\Omega^2 g^{-1}\  \sin\theta\cos\theta ,
    \label{7}
 \end{equation}
where $q$ is the rotation inhomogeneity parameter (\ref{1}).
%%%%%%%%%%%%%%%%%%%%%%%%%%%%%%%%%%%%%%%%%%%%%%%%%%%%%%%%%%%%%%%%%%%
 \bll
 \centerline{\bf Linear Stability Equations}
 \bl
%%%%%%%%%%%%%%%%%%%%%%%%%%%%%%%%%%%%%%%%%%%%%%%%%%%%%%%%%%%%%%%%%%%
The main approximations and methods of deriving
the equations for small disturbances were discussed
in detail previously (Kitchatinov 2008; Kitchatinov
\& R\"udiger 2008). This paper differs only in
allowance for the deviation of stratification from barotropy.
Therefore, the equations of the linear stability
problem will be written without repeating their
derivation. We repeat, however, the main
approximations and assumptions used in deriving these equations.

The initial equilibrium state does not depend on
time and longitude. Therefore, the dependence of the disturbances on
longitude and time in the linear stability
problem can be written as $\mathrm{exp}(\mathrm{i}m\phi - \mathrm{i}\omega t)$, where $m$ is the azimuthal wave
number. A positive imaginary part
of the eigenvalue, $\Im (\omega ) > 0$, means an instability.

Stable stratification of the radiation zone prevents
mixing on large radial scales. Therefore, the radial
scale of disturbances is assumed to be small and the
stability analysis is local in radius: perturbations of the velocity, $\vec u$, and entropy, $s'$, depend on radius
as $\mathrm{exp}(\mathrm{i}kr)$ with $kr \gg 1$. At the same time, the
mixing in horizontal directions encounters
no counteraction and the stability analysis is global
in these directions. As we will see, the most unstable
disturbances actually have large horizontal scales.

We use the incompressibility approximation, $\mathrm{div}{\vec u} = 0$. It is justified for disturbances whose wavelength
in the radial direction is small compared to the
pressure scale height. The magnetic fields are
disregarded. The angular velocity is assumed to be
dependent on radius only but not on latitude.

The equations are written for the scalar potentials $P_u$ and $T_u$ of the the poloidal and toroidal components of the velocity perturbations:
 \begin{equation}
    {\vec u} = \frac{{\vec e}_r}{r^2}\hat{L}P_u
    - \frac{{\vec e}_\theta}{r}\left(\frac{\mathrm{i}m}{\sin\theta} T_u + \mathrm{i}k \frac{\partial P_u}{\partial\theta}\right)
    + \frac{{\vec e}_\phi}{r}\left(\frac{\partial T_u}{\partial\theta} + \frac{k m}{\sin\theta}P_u\right)
    \label{8}
 \end{equation}
(Chandrasekhar 1961), where
 \begin{equation}
    \hat{L} = \frac{1}{\sin\theta}\frac{\partial}{\partial\theta} \sin\theta\frac{\partial}{\partial\theta} - \frac{m^2}{\sin^2\theta}
    \label{9}
 \end{equation}
is the angular part of the Laplacian. We use non-dimensional
variables. The physical quantities can
be restored from the normalized perturbations of
entropy ($S$) and the poloidal ($V$) and the toroidal ($W$) flow
potentials using Eq.\,(\ref{8}) and the relations
 \begin{equation}
    s' = -\frac{\mathrm{i}c_\mathrm{p}N^2}{gk} S,\ \
    P_u = \left(\Omega r^2/k\right) V,\ \
    T_u = \Omega r^2 W .
    \label{10}
 \end{equation}

The equation for the entropy perturbations is
 \begin{equation}
    \hat{\omega} S =
    -\mathrm{i}\frac{\epsilon_\chi}{\hat{\lambda}^2} S
    + \hat{L}V
    + \mathrm{i}\frac{Q}{\hat{\lambda}}\mu\left( mW - (1-\mu^2)\frac{\partial V}{\partial\mu}\right) ,
    \label{11}
 \end{equation}
where $\hat{\omega} = \omega/\Omega - m$ is the dimensionless eigenvalue
in the co-rotating frame of reference, $\mu = \cos\theta$, $\hat{\lambda}$ and $Q$ are the two basic parameters controlling the influence of fluid stratification and differential rotation
 \begin{equation}
    \hat{\lambda} = \frac{N}{\Omega kr},\ \ \ Q = 2 q\frac{\Omega}{N}.
    \label{12}
 \end{equation}
The finite diffusion is taken into account in the parameters
 \begin{equation}
    \epsilon_\chi = \frac{\chi N^2}{\Omega^3 r^2} ,\ \ \
    \epsilon_\nu = \frac{\nu N^2}{\Omega^3 r^2} ,
    \label{13}
 \end{equation}
where $\chi$ and $\nu$ are the thermal diffusivity and viscosity,
respectively.

The complete system consists of three equations. In
addition to Eq.\,(\ref{11}) for the entropy perturbations, it
includes the equations for the poloidal flow,
 \begin{equation}
   \hat{\omega}(\hat{L}V) =  -\mathrm{i}\frac{\epsilon_\nu}{\hat{\lambda}^2}(\hat{L}V)
   - \hat{\lambda}^2(\hat{L}S)
   + 2mV - 2\mu (\hat{L}W) - 2(1-\mu^2)\frac{\partial W}{\partial\mu} ,
   \label{14}
 \end{equation}
and the toroidal flow,
 \begin{equation}
   \hat{\omega}(\hat{L}W) =  -\mathrm{i}\frac{\epsilon_\nu}{\hat{\lambda}^2}(\hat{L}W)
   + 2mW - 2\mu (\hat{L}V) - 2(1-\mu^2)\frac{\partial V}{\partial\mu}.
   \label{15}
 \end{equation}
The eigenvalue problem for the system of equations
(\ref{11}), (\ref{14}), and (\ref{15}) was solved numerically. The independent variables were expanded in a series of the associated Legendre polynomials, for
example,
 \begin{equation}
    S = \sum\limits_{l=\max(\mid m\mid , 1)}^K S_l P_l^{\mid m\mid }(\mu ) ,
    \label{16}
 \end{equation}
and similarly for $W$ and $V$.  This leads to a system of linear algebraic equations for the expansion amplitudes $S_l$, $W_l$ and $V_l$. The number of equations in the system is not about $3K$ but a factor of 2 smaller, because
the complete system splits into two independent
subsystems governing the eigenmodes symmetric and antisymmetric
relative to the equator.

Most of the calculations were performed for the following values of dissipation
parameters of Eq.\,(\ref{13}),
 \begin{equation}
    \epsilon_\chi = 10^{-4},\ \ \epsilon_\nu = 2\times 10^{-10} ,
    \label{17}
 \end{equation}
typical of the upper part of the solar radiation zone
(Kitchatinov \& R\"udiger 2008). In the cases where
we used other values, this is stipulated.
%%%%%%%%%%%%%%%%%%%%%%%%%%%%%%%%%%%%%%%%%%%%%%%%%%%%%%%%%%%%%%%%%%%
 \bll
 \centerline{\bf Symmetry Properties}
 \bl
%%%%%%%%%%%%%%%%%%%%%%%%%%%%%%%%%%%%%%%%%%%%%%%%%%%%%%%%%%%%%%%%%%%
Two types of equatorial symmetry are possible: symmetric modes for which $S(\mu ) = S(-\mu ),\ V(\mu ) = V(-\mu )$ and $W(\mu ) = -W(-\mu )$, and antisymmetric
modes with symmetric $W$ and antisymmetric $S$ and $V$. For the symmetric and antisymmetric modes, we will use the notations S$m$ and A$m$, respectively,
where $m$ is the azimuthal wave number. These notations correspond to the symmetry relative to the mirror-reflection about the equatorial plane. For example, for the S$m$-modes, $u_r$ and $u_\phi$ are symmetric relative to the equator, while $u_\theta$ is antisymmetric.

A more significant property of the system of equations
(\ref{11}), (\ref{14}), and (\ref{15}) consists in its symmetry
relative to the transformation
 \begin{equation}
    (q,m,\hat\omega ,W,V,S) \rightarrow (-q,-m,-\hat\omega^*,-W^*,V^*,-S^*) ,
    \label{18}
 \end{equation}
where the asterisk denotes complex conjugation.
This means that if the mode with some $m$ is unstable
at a certain rotation inhomogeneity $q$, then at a
rotation inhomogeneity of opposite sense ($-q$) there
is an unstable mode with the same growth rate and
azimuthal wave number $-m$. Therefore, it will suffice
to consider the stability, for example, only for $q > 0$;
the stability properties for $q < 0$ will then be known.
Below, we consider only the case where the rotation
rate increases with depth, i.e., $q > 0$.

Transformation (\ref{18}) also shows that stability
properties depend on the sign of the azimuthal
wave number $m$. This dependence usually implies
that unstable modes possess a finite helicity (R\"udiger
et al. 2012). The absolute helicity in the linear problem
is indefinite, but the relative helicity
 \begin{equation}
    H_\mathrm{rel} = \langle {\vec u}\cdot ({\vec\nabla}\times{\vec u})\rangle /(k\overline{u^2}) ,
    \label{19}
 \end{equation}
can be defined\footnote{Only the real components of the disturbances
are used to calculate the relative helicity (19), along with
any other nonlinear characteristics of the disturbances. The
imaginary parts are omitted.}. The angular brackets here denote
the azimuthal averaging:
 \begin{equation}
    \langle X\rangle = \frac{1}{2\pi}\int\limits_0^{2\pi} X \mathrm{d}\phi .
    \label{20}
 \end{equation}
For axisymmetric modes ($m = 0$), this corresponds
to averaging over the oscillation phase $\phi$ (the linear
solutions are determined to within phase factor $\mathrm{e}^{\mathrm{i}\phi}$). The overline in (\ref{19}) and below denotes averaging
over a spherical surface:
 \begin{equation}
    \overline{u^2} = \frac{1}{2}\int\limits_{-1}^1\langle u^2\rangle \mathrm{d}\mu .
    \label{21}
 \end{equation}

For barotropic fluids, the total (volume-integrated)
kinetic helicity is an integral of motion. For baroclinic
fluids, this is not the case. As we will see, the unstable
modes of baroclinic instability are indeed
helical.
%%%%%%%%%%%%%%%%%%%%%%%%%%%%%%%%%%%%%%%%%%%%%%%%%%%%%%%%%%%%%%%%%%%
 \newpage
 \centerline{\bf Two Modes of Stable Oscillations}
 \bl
%%%%%%%%%%%%%%%%%%%%%%%%%%%%%%%%%%%%%%%%%%%%%%%%%%%%%%%%%%%%%%%%%%%
The solutions for special limiting cases are helpful in discussing the results to follow. In this Section,
we consider uniform rotation ($q = 0$) in the absence
of dissipation ($\chi = \nu = 0$).

In the limiting case of a \lq\lq very stable'' stratification, $\hat\lambda \gg 1$, or $N \gg \Omega kr$, two modes of stable oscillations can be revealed:

(1) For one of them, the frequency is low, $\hat\omega \ll \hat\lambda$.
It then follows from Eq.\,(\ref{14}) that $S = 0$ and Eq.\,(\ref{11}) gives $V = 0$. The flow possesses no poloidal component and the spectrum of purely toroidal oscillations can be found from Eq.\,(\ref{15}):
 \begin{equation}
    \hat{\omega} = - \frac{2m}{l(l+1)} .
    \label{22}
 \end{equation}
These are the $r$-modes of global oscillations also
known as Rossby waves.

(2) There is another solution for which the frequency
is not low, $\hat\omega \sim \hat\lambda$. In this case, Eq.\,(\ref{15}) in the highest order in $\hat\lambda$ gives $W=0$. The flow does not contain
a toroidal part. We write Eqs. (\ref{14})
and (\ref{11}), also in the highest order in $\hat\lambda$, as $\hat\omega(\hat{L}V) = -\hat{\lambda}^2(\hat{L}S)$ and $\hat{\omega}S = \hat{L}V$, respectively. Poloidal oscillations
with the following spectrum are found:
 \begin{equation}
    \hat{\omega} = \pm\hat{\lambda}\sqrt{l(l+1)}.
    \label{23}
 \end{equation}
As can be seen from the expression for the frequency
 \begin{equation}
    \omega = \pm\frac{N}{kr}\sqrt{l(l+1)} ,
    \label{24}
 \end{equation}
rotation does not affect these high-frequency oscillations.
These are the internal gravity waves or $g$-modes.

As we will see, in a differentially rotating fluid
with baroclinic stratification, both modes of global
oscillations acquire positive growth rates, i.e., become
unstable.
%%%%%%%%%%%%%%%%%%%%%%%%%%%%%%%%%%%%%%%%%%%%%%%%%%%%%%%%%%%%%%%%%%%
 \bll
 \centerline{\bf RESULTS AND DISCUSSION}
 \bl
%%%%%%%%%%%%%%%%%%%%%%%%%%%%%%%%%%%%%%%%%%%%%%%%%%%%%%%%%%%%%%%%%%%
 \centerline{\bf Stability Borders and Growth Rates}
 \bl
%%%%%%%%%%%%%%%%%%%%%%%%%%%%%%%%%%%%%%%%%%%%%%%%%%%%%%%%%%%%%%%%%%%
The lines separating the regions of stability
and instability for disturbances with different
equatorial and axial symmetries are shown in Fig.\,2.
The instability appears at a small rotation inhomogeneity.
In the upper part of the solar radiation zone, $N/\Omega \approx 400$. Even a weakly inhomogeneous rotation $q \sim 10^{-4}$ is unstable. In this way the instability under
consideration differs from the barotropic instabilities
that appear at a relatively large rotation inhomogeneity.
Another important difference is that baroclinic
instability exists both for axisymmetric disturbances
and for various azimuthal wave numbers $m\neq 0$. However, the
greater the number |m|, the larger rotation inhomogeneity
is required for the onset of instability.
This trend is confirmed by our calculations for $\mid m\mid \leq 10$. The disturbances that are global in horizontal dimensions
are most unstable.

\begin{figure}[htb]
 \centerline{
 \includegraphics[width=12 cm]{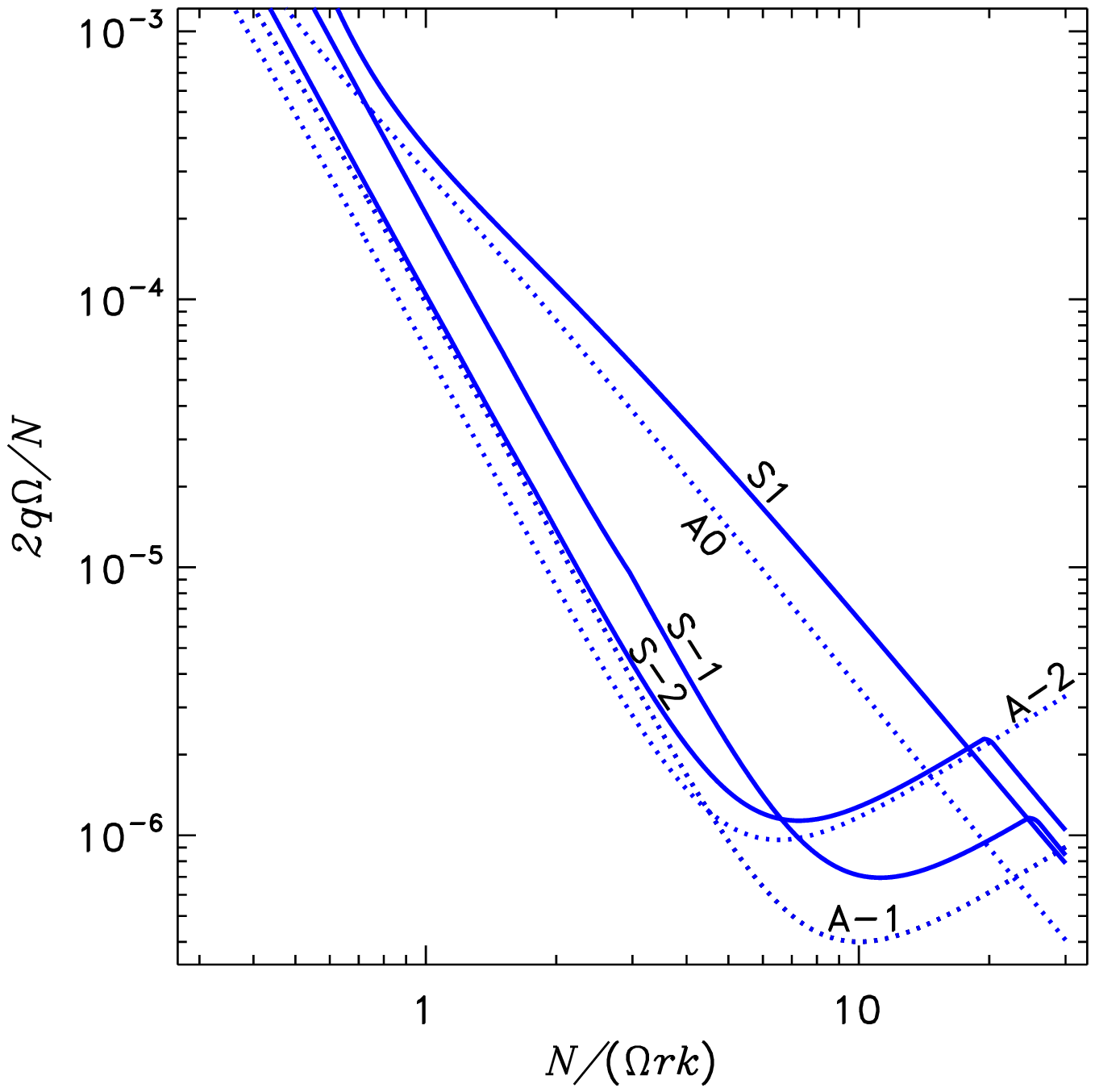}}
 \begin{description}
 \item{\small Fig.~2. Lines of neutral stability for symmetric
    (solid lines) and antisymmetric (dotted lines) disturbances about
    the equator. The lines are marked by the corresponding symmetry notations. The instability regions are above
    the lines.
    }
 \end{description}
\end{figure}

The lines for modes S-1 and S-2 in Fig.\,2 have
kinks. This implies that different line segments correspond
to disturbances of different nature. Unstable
disturbances close to the $r$- and $g$-modes of global
oscillations are revealed. This can be seen from the Table, where the characteristics of unstable disturbances
are given. The kinetic energy of the disturbances
is the sum of the energies of their poloidal and
toroidal components (Chandrasekhar 1961):
 \begin{equation}
    \overline{u^2} = \overline{u^2_\mathrm{p}} + \overline{u^2_\mathrm{t}} = \frac{1}{4}\sum\limits_l l(l+1) \left( \mid V_l\mid^2 + \mid W_l\mid^2\right) .
    \label{25}
 \end{equation}
The Table lists the closest frequencies of the poloidal
$g$-modes (\ref{23}) for unstable poloidal disturbances ($\overline{u^2_\mathrm{p}}/\overline{u^2_\mathrm{t}} > 1$) and the closest frequencies of the toroidal
$r$-modes (\ref{22}) for toroidal disturbances ($\overline{u^2_\mathrm{p}}/\overline{u^2_\mathrm{t}} < 1$). The frequencies of the unstable disturbances and the
corresponding oscillation modes differ little; there is
a correspondence to the largest-scale oscillations.
For example, the frequency of the unstable poloidal
disturbance A3 is 13.6. The lowest value of $l$
in expansion (16) for the poloidal potential of this
disturbance is $l = 4$. For $l = 4$ we find a frequency of 13.4
close to that of the unstable disturbance from (\ref{23}).
The cases where there is no correspondence to the
largest-scale oscillation mode are marked with an
asterisk in the Table. For example, the expansion of
the toroidal potential for the toroidal mode A-3 with a
frequency of 0.199 begins from $l = 3$, but the $r$-mode
(\ref{22}) with the next $l = 5$ has the closest frequency
$\hat{\omega}^r = 0.2$ (the summation in (\ref{16}) for disturbances
with a certain equatorial symmetry is over either even
or odd $l$). The correspondence of the frequencies
and the poloidal or toroidal character of growing disturbances to
the $r$- and $g$-modes of global oscillations allows us to interpret the
instability as the
loss of stability against the excitation of these global
oscillations. The question of what determines the transport
of angular momentum in differentially rotating
stars, the instability or the $g$-modes
(Spruit 1987; Charbonnel \& Talon 2005), may find
an unexpected answer: the instability excites the $g$-modes.

The Table also gives the correlation of the entropy and radial velocity
perturbations, to which the power supplied by buoyancy forces is proportional. This correlation is positive for all unstable
modes. Calculations show that this correlation
can be negative only for damped disturbances. The
energy of the growing disturbances increases due to
the work of buoyancy forces, as it should be for
baroclinic instability (Fig.\,1).

\begin{table}[htb]
 \caption{Parameters of unstable disturbances for $\hat{\lambda} = 3$ and $Q=10^{-3}$: $\hat{\gamma}$ is the disturbance growth rate (26), $\Re (\hat{\omega})$ is the oscillation
frequency, $\hat{\omega}^r$ and $\hat{\omega}^g$ are the closest frequencies of the $r$- or $g$-modes (\ref{22}) or (\ref{23}), respectively, $\overline{u^2_\mathrm{p}}/\overline{u^2_\mathrm{t}}$ is the ratio of the
energies of the poloidal and toroidal flow components, and $\overline{Su_r}/\sqrt{\overline{u^2_r}\ \overline{S^2}}$ is the relative correlation of the entropy and radial velocity perturbations.}
 \vspace{0.2 truecm}
 \centerline{
 \begin{tabular}{ccccccc}
 \hline
 Mode & $\hat{\gamma}, 10^{-4}$ & $\Re (\hat{\omega})$ & $\hat{\omega}^r$ & $\hat{\omega}^g$ & $\overline{u^2_\mathrm{p}}/\overline{u^2_\mathrm{t}}$ & $\overline{Su_r}/\sqrt{\overline{u^2_r}\ \overline{S^2}}$ \\
 \hline
 A0 & $8.41$ & $4.34$ &    & $4.24$  & $24.0$ & $3.35\times 10^{-5}$ \\
 A1 & $2.91$ & $7.27$ &    & $7.35$  & $43.0$ & $6.38\times 10^{-6}$ \\
 A3 & $1.14$ & $-13.6$ &    & $-13.4$  & $171$ & $1.38\times 10^{-6}$ \\
 A10& $0.709$ & $-34.6$ &    & $-34.5$  & $2487$ & $3.32\times 10^{-7}$ \\
 \hline
 A-1 & $2.04$ & $0.989$ &  $1$  &   & $1.99\times 10^{-4}$ & $2.99\times 10^{-3}$ \\
 A-3 & $1.46$ & $0.199$ &  $0.2^*$  &   & $9.62\times 10^{-7}$ & $2.17\times 10^{-2}$ \\
 A-10 & $0.507$ & $0.0952$ &  $0.0952^*$  &   & $2.79\times 10^{-9}$ & $0.155$ \\
 \hline
 S0 & $3.29$ & $7.46$ &    & $7.35$  & $32.7$ & $7.46\times 10^{-6}$ \\
 S1 & $4.93$ & $-4.85$ &    & $-4.24$  & $34.6$ & $2.17\times 10^{-5}$ \\
 S3 & $3.10$ & $-10.7$ &    & $-10.4$  & $260$ & $4.90\times 10^{-6}$ \\
 S10 & $1.03$ & $-31.5$ &    & $-31.5$  & $5727$ & $5.24\times 10^{-7}$ \\
 \hline
 S-1 & $2.94$ & $0.320$ &  $0.333$  &  & $1.93\times 10^{-4}$ & $3.67\times 10^{-3}$ \\
 S-3 & $2.22$ & $-10.2$ &    & $-10.4$  & $265$ & $3.35\times 10^{-6}$ \\
 S-10 & $0.874$ & $-31.4$ &    & $-31.5$  & $5772$ & $4.44\times 10^{-7}$ \\
 \hline
 \end{tabular}
 }

$^*$The asterisk marks the frequencies that do not correspond to the largest-scale oscillations, i.e., not to the smallest $l$ in Eq.\,(\ref{22}) for a
given mode.
 \label{table}
 \end{table}

The Table gives the disturbance growth rate
 \begin{equation}
    \hat{\gamma} = 2\pi\Im (\hat{\omega}) ,
    \label{26}
 \end{equation}
normalized to the rotation period, i.e., the disturbances
grow by a factor of $\mathrm{e}^{\hat{\gamma}}$ in one stellar rotation. The growth rates are small. The star makes about 10\,000 rotations in a disturbance e-folding time. Even for slowly rotating stars, however, this time ($\sim 1000$ yr) is short compared to evolutionary time scales. In Fig.\,3, the growth rate of disturbances
is plotted against their dimensionless wavelength $\hat{\lambda}$ (\ref{12}). The equatorial symmetry does not
determine the properties of unstable modes uniquely. For example, there is a discrete spectrum of modes S1. Figure\,3 shows the highest growth
rates. The kinks in the lines for modes with negative
$m$ correspond to a change in the type of the most
rapidly growing disturbance. The highest growth
rates belong to the $r$-modes for relatively small $\hat\lambda$ and
to the $g$-modes for large $\hat\lambda$.

\begin{figure}[htb]
 \centerline{
 \includegraphics[width=12 cm]{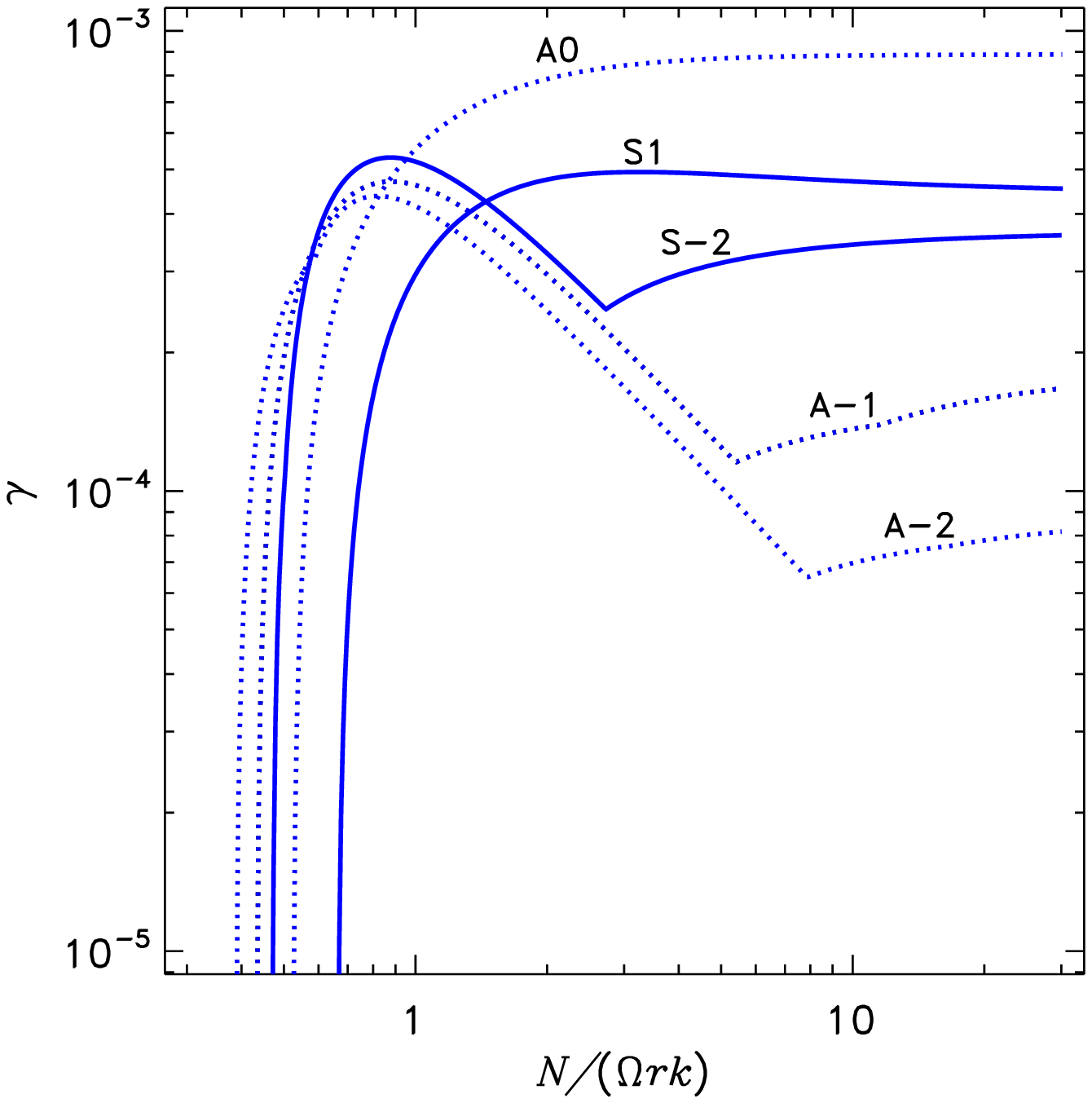}}
 \begin{description}
 \item{\small Fig.~3. Growth rates (\ref{26}) of unstable disturbances versus their radial
    wavelength $\hat\lambda$ for a rotation inhomogeneity parameter
    $Q = 0.001$. The solid and dotted lines show the results
    for the modes symmetric and antisymmetric about
    the equator, respectively.
    }
 \end{description}
\end{figure}

\begin{figure}[htb]
 \centerline{
 \includegraphics[width=12 cm]{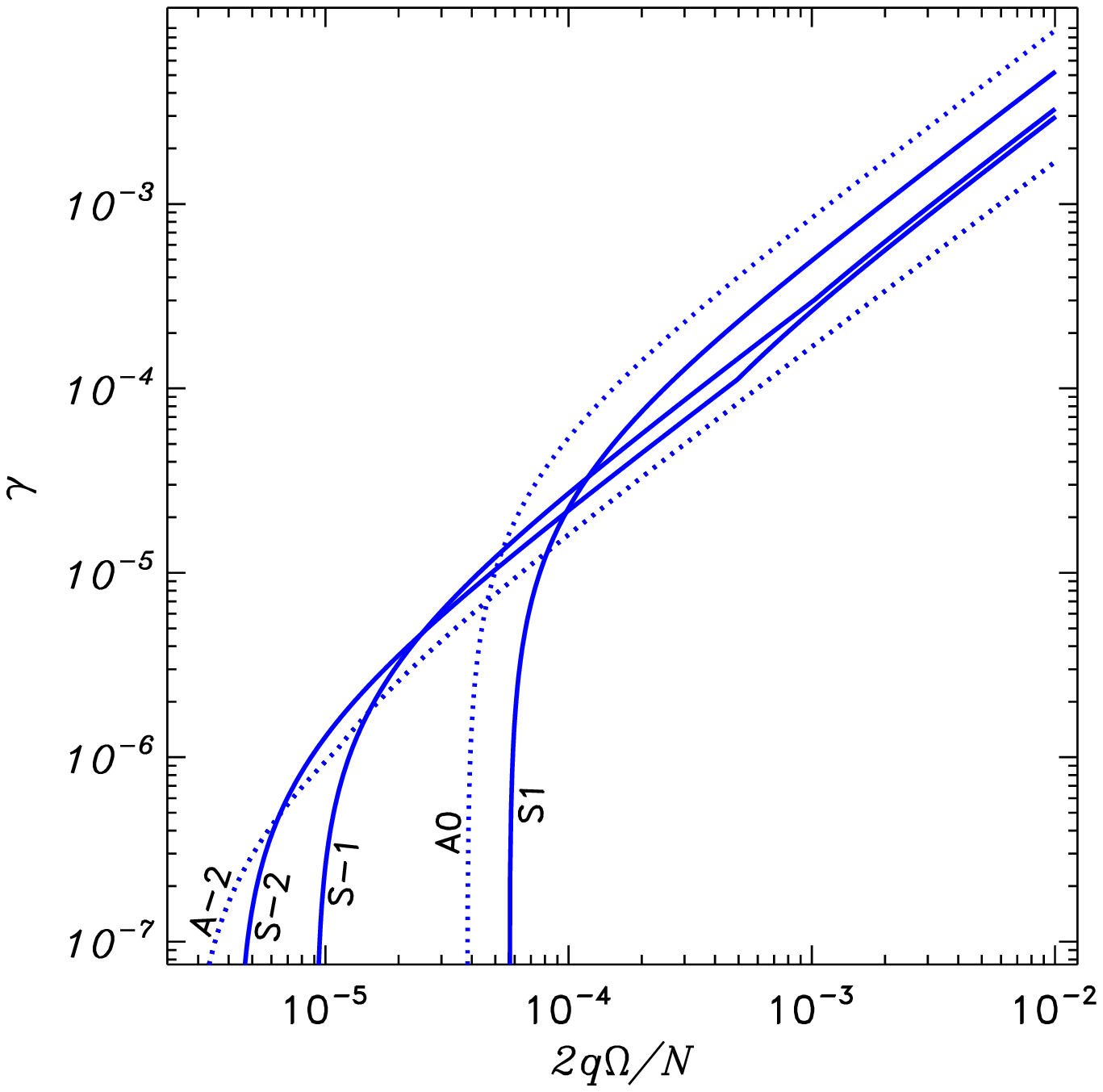}}
 \begin{description}
 \item{\small Fig.~4. Growth rates (\ref{26}) of unstable disturbances versus rotation
    inhomogeneity parameter $Q$ (\ref{12}) for $\hat{\lambda} = 3$.
    }
 \end{description}
\end{figure}

In Fig.\,4, the highest growth rates are plotted
against the rotation inhomogeneity parameter $Q$ (\ref{12}).
For relatively large $Q$, these dependencies are nearly
linear, $\hat{\gamma} \sim Q$.
%%%%%%%%%%%%%%%%%%%%%%%%%%%%%%%%%%%%%%%%%%%%%%%%%%%%%%%%%%%%%%%%%%%
 \bll
 \centerline{\bf Dependence on Thermal Conductivity}
 \bl
%%%%%%%%%%%%%%%%%%%%%%%%%%%%%%%%%%%%%%%%%%%%%%%%%%%%%%%%%%%%%%%%%%%
The dependence on thermal conductivity is of interest
in connection with the possible influence of
chemical composition inhomogeneity. Such inhomogeneity
is important for stability. The
increase in mean molecular weight $\mu$ with depth makes the stratification \lq\lq more stable''. This
can be taken into account by replacing the frequency
$N$ (\ref{4}) with its effective value $N_*$,
 \begin{equation}
    N_*^2 = N^2 + N_\mu^2, \ \ \ N_\mu^2 = -\frac{g}{\mu}\frac{\mathrm{d}\mu}{\mathrm{d} r}
    \label{27}
 \end{equation}
(Kippenhahn \& Weigert 1990). This, however, is not the only effect of the
compositional gradient. The diffusivity for chemical inhomogeneities
in stellar radiation zones is much smaller
than the thermal diffusivity. Therefore, the inhomogeneity
of $\mu$ reduces the dissipation rate of density
inhomogeneities in unstable disturbances.

Here, we do not account for composition inhomogeneity,
but the character of its influence can be
seen by analyzing dependence of the instability on thermal conductivity.
Figure\,5 shows the growth rates of the
unstable $g$-mode A0 for three values of the dimensionless
thermal diffusivity $\epsilon_\chi$ (\ref{13}). Similar results are
also obtained for other unstable modes. An increase
in $\epsilon_\chi$ suppresses the instability.

\begin{figure}[htb]
 \centerline{
 \includegraphics[width=12 cm]{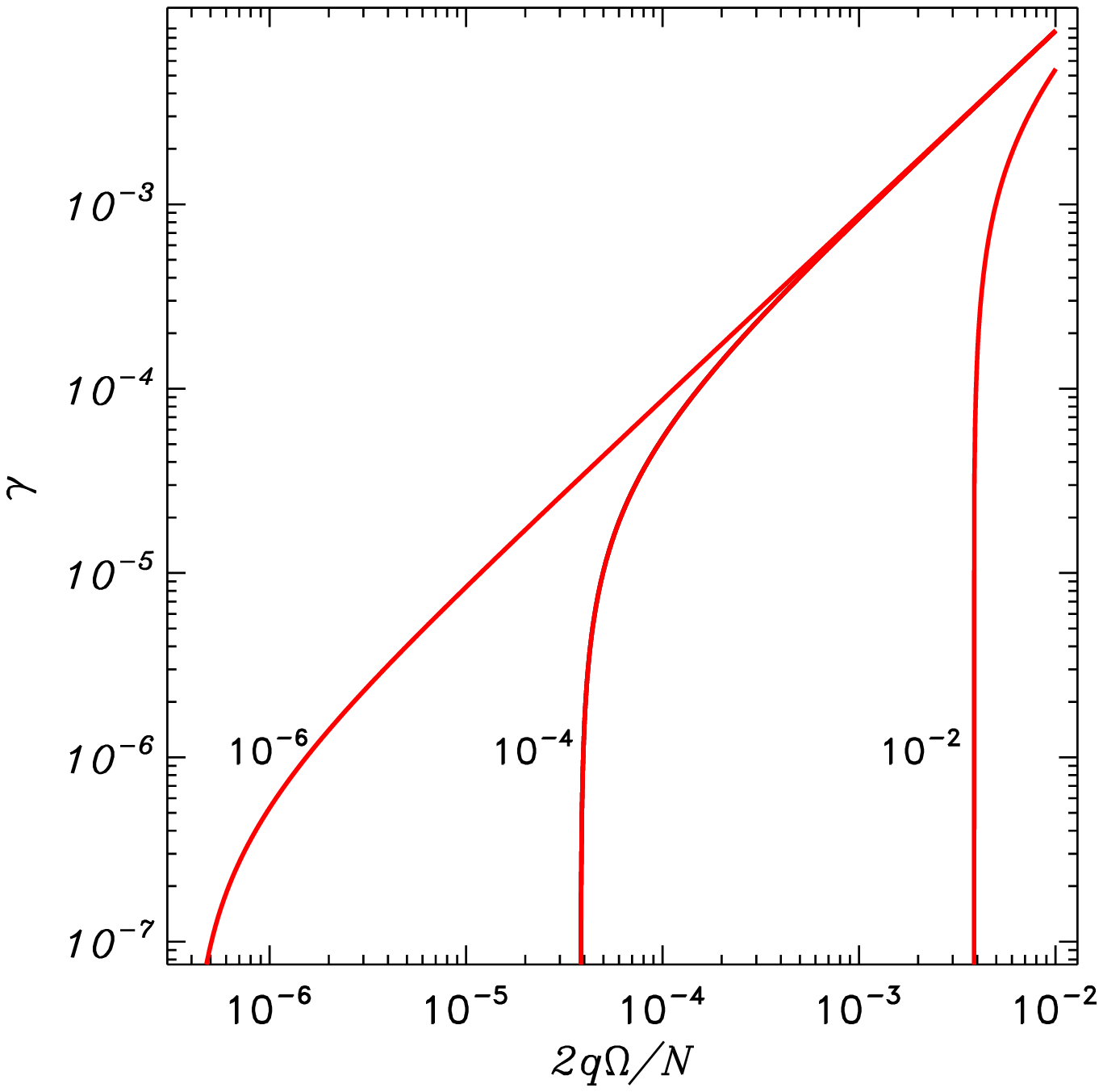}}
 \begin{description}
 \item{\small Fig.~5. Growth rate (\ref{26}) of the unstable mode A0 versus
        rotation inhomogeneity parameter Q (\ref{12}) for three values of the
        normalized thermal diffusivity $\epsilon_\chi$ (\ref{13}) for $\hat{\lambda} = 3$. The curves are marked by the corresponding values of $\epsilon_\chi$.
    }
 \end{description}
\end{figure}

It is generally believed that conduction of heat
{\em amplifies} the instabilities in stellar radiation
zones. Radial displacements produce the temperature and density disturbances and are, therefore, opposed by buoyancy.
The dissipation of temperature inhomogeneities
reduces the stabilizing buoyancy effect, thereby amplifying the instabilities.

Figure\,5 shows that the opposite is true of the
baroclinic instability. This instability is peculiar in
that it emerges precisely due to special features of
the radiation zone stratification and is produced by
buoyancy forces (Fig.\,1). Therefore, an increase in
thermal conductivity suppresses this instability.
Assertions in the literature that the compositional gradient in stellar radiation zones switches off baroclinic instability seem questionable.
%%%%%%%%%%%%%%%%%%%%%%%%%%%%%%%%%%%%%%%%%%%%%%%%%%%%%%%%%%%%%%%%%%%
 \newpage
 \centerline{\bf Helicity and the Possibility of Dynamo}
 \bl
%%%%%%%%%%%%%%%%%%%%%%%%%%%%%%%%%%%%%%%%%%%%%%%%%%%%%%%%%%%%%%%%%%%
Figure\,6 shows the distributions of the relative
helicity (\ref{19}) for three $g$-modes of the instability under
consideration. Positive and negative helicities
dominate in the northern and southern hemispheres,
respectively. In the dynamo theory, the helicity is
known to be important for magnetic field generation.

\begin{figure}[htb]
 \centerline{
 \includegraphics[width=12 cm]{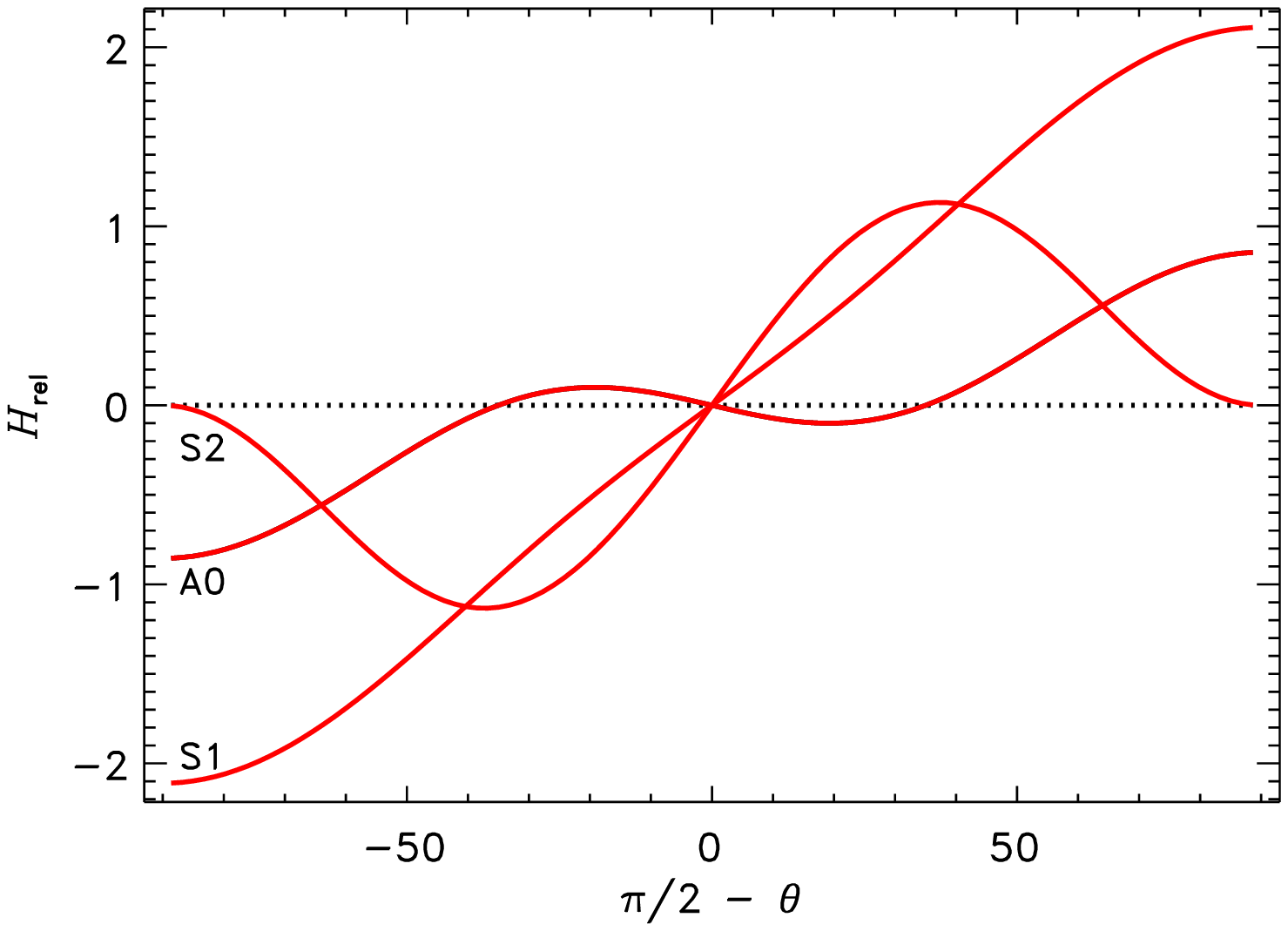}}
 \begin{description}
 \item{\small Fig.~6. Relative helicity (\ref{19}) versus latitude for the three
    most rapidly growing instability modes ($\hat{\lambda} = 3,\ Q = 0.001$).
    }
 \end{description}
\end{figure}

The origin of magnetic fields in stellar radiation
zones presents a problem. Solar-type stars at early
evolutionary stages are fully convective for more than
a million years, which is approximately a factor of $10^4$
longer than the turbulent diffusion time. A hydromagnetic
dynamo can operate in such fully convective
stars (Dudorov et al. 1989). Subsequently, a radiative
core emerges and grows in the central part of the star.
During its growth, it can capture the magnetic field
from the surrounding convective envelope. However,
this field is weak (<1\,G), because the convective
dynamo field is oscillating and the frequency of its
oscillations is much higher than the growth rate of the
radiation zone (Kitchatinov et al. 2001). The helicity
of the eigenmodes of baroclinic instability (Fig.\,6)
points to another possibility — the dynamo action in
a differentially rotating unstable radiation zone.

The radiation zones of solar-type stars are deep
beneath the surface and are inaccessible to direct
observations. Higher-mass stars have outer radiative
envelopes. Differential rotation can be present in
such stars as they approach the main sequence due
to radially nonuniform contraction. Recently, Alecian
et al. (2013) detected rapid (in several years) changes
of the global magnetic field on one of such Herbig
Ae/Be stars with an extended outer radiation
zone. They interpreted these changes as a manifestation
of a deep dynamo in the newly-born convective
core. However, an alternative explanation is also possible — the dynamo action due to baroclinic instability
in a differentially rotating radiative envelope.
%%%%%%%%%%%%%%%%%%%%%%%%%%%%%%%%%%%%%%%%%%%%%%%%%%%%%%%%%%%%%%%%%%%
 \bll
 \centerline{\bf CONCLUDING REMARKS}
 \bl
%%%%%%%%%%%%%%%%%%%%%%%%%%%%%%%%%%%%%%%%%%%%%%%%%%%%%%%%%%%%%%%%%%%
Linear analysis does not permit determination of the final state to which instability growth will lead.
However, one might expect fully developed
turbulence in view of the great variety of baroclinic
instability modes. Turbulence in the radiation
zone, irrespective of its source, is highly anisotropic
with a predominance of horizontal flows, $\overline{u_r^2}/\overline{u^2} \sim \Omega^2/(\tau^2N^4) \ll 1$, where $\tau$ is the eddy turnover time
(Kitchatinov \& Brandenburg 2012). Such turbulence
efficiently transports angular momentum, removing
rotation inhomogeneity. Note that the
transport of angular momentum by anisotropic turbulence
is not reduced to the action of eddy viscosity
(Lebedinskii 1941). There are nondissipative angular
momentum flows; as a result, the smoothing of
rotation inhomogeneities in stellar radiation zones is
much faster than the diffusion of chemical species. Since the threshold value of differential
rotation for the onset of baroclinic instability is
very low (Fig.\,2), this instability can lead to an essentially
uniform rotation of the radiation zone, which is
revealed by helioseismology.

Baroclinic instability can also have a bearing
on the origin of magnetic fields in stellar radiation
zones. The convective instability in rotating stars is
known to be capable of generating magnetic fields.
The helicity of convective flows plays the most important
role in this process (see, e.g., Vainshtein
et al. 1980). The growing global modes of baroclinic instability also possess helicity and may be capable of
generating magnetic fields.

Baroclinic instability has a bearing not only on
stars. Already Tassoul \& Tassoul (1983)
pointed to this instability as a possible cause of
turbulence in accretion disks. Subsequently, Klahr
\& Bodenhaimer (2003) analyzed this possibility.
The so-called stratorotational instability of a Couette
flow (Shalybkov \& R\"udiger 2005) is also most likely
of the baroclinic type.

Figure\,2 shows that the threshold rotation inhomogeneity
for the onset of instability decreases
with increasing radial scale of the $g$-modes.
Therefore, a stability analysis for disturbances
that are global not only horizontally but also
radially can be a perspective for further study of
baroclinic instability.
%%%%%%%%%%%%%%%%%%%%%%%%%%%%%%%%%%%%%%%%%%%%%%%%%%%%%%%%%%%%%%%%%%%
 \bll

{\bf Acknowledgements.} This work was supported by the Russian Foundation
for Basic Research (project no. 12-02-92691\_Ind)
and the Ministry of Education and Science of the
Russian Federation (contract 8407 and State contract
14.518.11.7047).
%%%%%%%%%%%%%%%%%%%%%%%%%%%%%%%%%%%%%%%%%%%%%%%%%%%%%%%%%%%%%%%%%%%
\newpage
%%%%%%%%%%%%%%%%%%%%%%%%%%%%%%%%%%%%%%%%%%%%%%%%%%%%%%%%%%%%%%%%%%%

\centerline{\bf REFERENCES}
\begin{description}
\item Acheson,~D.\,J.
    1978, Phyl. Trans. Roy. Soc. London {\bf A289}, 459
\item  Alecian,~E., Neiner,~C., Mathis,~S. et al.
    2013, \aa\ {\bf 549}, L8
\item Barnes,~S.\,A.
    2003, \apj\ {\bf 586}, 464
\item Chandrasekhar,~S.
    1961, {\sl Hydrodynamic and Hydromagnetic Stability,} Oxford, Clarendon Press, p.622
\item Charbonneau,~P., Dikpati,~M., \& Gilman,~P.\,A.
    1999, \apj\ {\bf 526}, 523
\item Charbonnel,~C., \& Talon,~S.
    2005, Science {\bf 309}, 2189
\item Denissenkov,~P.\,A., Pinsonneault,~M., Terndrup,~D.\,M., \& Newsham,~G.
    2010, \apj\ {\bf 716}, 1269
\item Dicke,~R.\,H.
    1970, \araa\ {\bf 8}, 297
\item Dudorov,~A.\,E., Krivodubskii,~V.\,N., Ruzmaikina,~T.\,V., \& Ruzmaikin,~A.\,A.
    1989, \aj\ {\bf 66}, 809
\item Gilman,~P.\,A., Dikpati,~M., \& Miesch,~M.\,S.
    2007, \apjs\ {\bf 170}, 203
\item Goldreich,~P, \& Schubert,~G.
    1967, \apj\ {\bf 150}, 571
\item Hartmann,~L.\,W., \& Noyes,~R.\,W.
    1987, \araa\ {\bf 25}, 271
\item Kippenhahn,~R., \& Weigert,~A.
    1990, {\sl Stellar Structure and Evolution,} Berlin, Springer
\item Kitchatinov,~L.\,L.
    2008, \aj\ {\bf 85}, 279
\item Kitchatinov,~L.\,L.
    2010, \aj\ {\bf 87}, 3
\item Kitchatinov,~L.\,L., \& Brandenburg,~A.
    2012, \an\ {\bf 333}, 230
\item Kitchatinov,~L.\,L., \& R\"udiger,~G.
    2008, \aa\ {\bf 478}, 1
\item Kitchatinov,~L.\,L., Jardine,~M., \& Collier Cameron,~A.
    2001, \aa\ {\bf 374}, 250
\item Klahr,~H., \& Bodenhaimer,~P.
    2003, \apj\ {\bf 582}, 869
\item Korycansky,~D.\,G.
    1991, \apj\ {\bf 381}, 515
\item Lebedinskii,~A.\,I.
    1941, {\sl Astron. Zh.} {\bf 18}, 10
\item Pinsonneault,~M.
    1997, \araa\ {\bf 35}, 557
\item R\"udiger,~G., Kitchatinov,~L.\,L., \& Elsther,~D.
    2012, \mnras\ {\bf 425}, 2267
\item Shalybkov,~D., \& R\"udiger,~G.
    2005, \aa\ {\bf 438}, 411
\item Shibahashi,~H.
    1980 {\sl PASJ} {\bf 32}, 341
\item Shou,~J., Antia,~H.\,M., Basu,~S. et al.
    1998, \apj\ {\bf 505}, 390
\item Skumanich,~A.
    1972, \apj\ {\bf 171}, 565
\item Spruit,~H.\,C.
    1987, {\sl The Internal Solar Angular Velocity}, Ed. B.R.\,Durney, S.\,Sofia, Dordrecht: D.\,Reidel Publ., p.185
\item Spruit,~H.\,C., \& Knobloch,~E.
    1984, \aa\ {\bf 132}, 89
\item Tassoul,~J.-L.
    1982, {\sl Theory of Rotating Stars}, Princeton, Princeton Univ. Press
\item Tassoul,~M., \& Tassoul,~J.-L.
    1983, \apj\ {\bf 271}, 315
\item Vainshyein,~S.\,I., Zeldovich,~Ya.\,B., \& Ruzmaikin,~A.\,A.
    1980, {\sl Turbulent Dynamo in Astrophysics}, Moscow, Nauka (in Russian)
\end{description}
%\bll
%\centerline{\it Translated by V.~Astakhov}
\end{document}